\documentclass[aps,pra,twocolumn,superscriptaddress,showpacs,showkeys]{revtex4}
\usepackage{amsmath}
\usepackage{amsfonts}            
\usepackage{amssymb}    

\usepackage[mathcal,mathscr]{eucal}

\usepackage{mathrsfs}

\usepackage{eufrak}

\usepackage{graphicx}% Include figure files
\usepackage{epstopdf}% Include figure files

\usepackage{dcolumn}% Align table columns on decimal point
\usepackage{bm}% bold math 

\makeatletter
\def\@dotsep{4.5}
\makeatother

\begin{document}

%\title{Diffractive contribution to the refractive index for matter waves}

\title{Sensitive imaging of electromagnetic fields with paramagnetic polar molecules}

\author{Sergey V. Alyabyshev}

\affiliation{Department of Chemistry, University of British Columbia, Vancouver, BC V6T 1Z1, Canada}

\email{salyabyshev@gmail.com}

\author{Mikhail Lemeshko}

%\email{mikhail.lemeshko@gmail.com}

\affiliation{%
ITAMP, Harvard-Smithsonian Center for Astrophysics and Harvard Physics Department,  Cambridge, Massachusetts 02138, USA
}%

\email{mikhail.lemeshko@gmail.com}

\author{Roman V. Krems}

\affiliation{Department of Chemistry, University of British Columbia, Vancouver, BC V6T 1Z1, Canada }

\email{rkrems@chem.ubc.ca}

\date{\today}% It is always \today, today,

\begin{abstract}
We propose a method for sensitive parallel detection of low-frequency electromagnetic fields based on the fine structure interactions in paramagnetic polar molecules. Compared to the recently implemented scheme employing ultracold $^{87}$Rb atoms [B{\"o}hi \textit{et al.}, Appl. Phys. Lett.~\textbf{97}, 051101 (2010)], the technique based on molecules offers a 100-fold higher sensitivity, the possibility to measure both the electric and magnetic field components, and a probe of a wide range of frequencies from the dc limit to the THz regime. 
\end{abstract}

%\pacs{34.50.Cx, 34.20.-b, 34.90.+q, 32.60.+i, 33.90.+h, 33.15.Kr, 37.10.Vz, 37.10.Pq}% PACS, the Physics and Astronomy
%\keywords{quantum gases, cold and ultracold collisions, dipole-dipole interaction, induced-dipole interaction,  AC Stark effect, dynamic polarizability, far-off-resonant laser field} %Use showkeys class option if keyword
\maketitle

%Sensitive detection of electromagnetic fields is an outstanding task in a variety of applications. 

\section{Introduction}

Sensitive detection of weak electromagnetic fields is critical for many applications ranging from fundamental physics measurements~\cite{BrownPRL10} to biomagnetic imaging of the brain and heart~\cite{XiaAPL06,BisonOptEx03} to the detection of explosive materials~\cite{DelaneyReport03}. A tremendous progress in measuring magnetic fields has been recently achieved leading to the development of Hall effect sensors~\cite{RamsdenHallBook06}, SQUID sensors~\cite{HuberRSI08}, force sensors~\cite{PoggioNano10}, sensors based on microelectromechanical systems~\cite{GojdkaAPL11}, and NV centers in diamond~\cite{TaylorNatPhys08}, as well as atomic magnetometers~\cite{BudkerNatPhys07,DangAPL10}, making it possible to achieve the magnetic field sensitivity of  $0.1$~fT Hz$^{-1/2}$ and to detect the magnetic field of a single electron, with steps being taken towards the detection of the magnetic field of a single nuclear spin \cite{ZhaoNatNano11,SchaffryPRL11}. At the same time, the development of scanning capacitance microscopy~\cite{WilliamsAPL89}, scanning Kelvin probe~\cite{HenningJAP95}, and electric field-sensitive atomic force microscopy~\cite{SchonenbergerPRL90} advanced
 the techniques for measuring electric fields to the level of probing individual charges. An unprecedented accuracy of $10^{-6}$ electron charge was achieved with the use of single-electron transistors~\cite{DevoretNature00}.
%Nitrogen-vacancy centers in diamond allow for very local measurements sensitive to electric fields of single charges~\cite{DoldeNatPhys11} and magnetic fields of a single electron spin~\cite{TaylorNatPhys08}, with steps being made towards the detection of single nuclear spins \cite{ZhaoNatNano11, SchaffryPRL11}.
Yet, even with the sensitivity pushed to its fundamental limit, none of these methods allows for parallel measurements, i.e.\ imaging of the field amplitudes and phases at many spatial points at the same time.
Recently B\"ohi \textit{et al.} proposed to use ultracold atoms for sensitive parallel imaging of microwave fields with frequencies in the range $2.5 - 14$ GHz~\cite{BohiAPL10}. The method relies on measuring the phase difference between two hyperfine states of $^{87}$Rb, accumulated due to an interaction with the magnetic component of the microwave field. The acquired phase difference is proportional to the evolution time, the magnetic moment of the atom, and the magnetic field amplitude. However, long measurement times lead to image blurring due to the atomic motion and decoherence, resulting in a compromise between field sensitivity and spatial resolution. 

Although measuring the electric component of an ac field is several orders of magnitude more efficient than detecting the magnetic component~\footnote{The electric, $E$, and magnetic, $B$, field amplitudes of the electromagnetic field are related as $E=B c$ with $c$ the speed of light, rendering the phase difference accumulated due to an electric dipole transition two orders of magnitude larger than for the transition between the atomic hyperfine states (for a given interaction time $\tau$).}, atoms possess no permanent electric dipole moments, rendering the detection of the Zeeman shift  the only possible option. Here we describe a technique for  parallel, noninvasive, and complete (amplitudes and phases) imaging of electromagnetic fields with an ensemble of polar open-shell molecules, many of which have been successfully cooled and trapped in experiments~\cite{WeinsteinCaH,BasPRL05,CampbellPRL07,ShumanNature10,ShumanPRL09}. We show that the presence of permanent electric and magnetic dipole moments and the variety of molecular rotational constants allow for the detection of both electric and magnetic field components, in a wide range of frequencies from the dc limit through the radio and microwave to the THz frequency range. We show that measuring the electric component of an oscillating field must result in shorter measurement times compared to the atomic experiments, and consequently higher spatial resolution, which is mainly limited by the optical detection scheme and photon scattering rate.  

\section{Theory}
\subsection{$^2\Sigma$ molecules in magnetic and electric fields}

First,  consider $^2\Sigma$ molecules with a dipole moment $\mathbf{d}$, subject to known dc magnetic and electric fields, $\mathbf{B}_0$ and $\mathbf{E}_0$, both pointing along the laboratory $Z$ axis. The molecules are described by the following Hamiltonian:
\begin{equation}
\label{Hamil2Sigma}
	H= b_r \mathbf{N}^2 + \gamma \mathbf{N \cdot S} + g_S \mu_B \mathbf{B_0 \cdot S} - \mathbf{d \cdot E_0},
\end{equation}
where $\mathbf{N}$ and $\mathbf{S}$ are the rotational and spin angular momenta of the molecule, $b_r$ and $\gamma$ are the rotational and spin-rotation interaction constants, $\mu_B$ is the Bohr magneton, and $g_S=2.0023$. 
In the absence of external fields, the states of a $^2 \Sigma$ molecule are labeled by $\vert N, J, M \rangle$, where $\mathbf{J} = \mathbf{N} + \mathbf{S}$ is the total angular momentum and $M$ is the projection of $\mathbf{J}$ on the $Z$ axis. We use the same quantum numbers to label the molecular states in the presence of dc fields, bearing in mind that $J$ is not conserved. 

\begin{figure}[t]
\includegraphics[width=7.1cm, trim = 70 0 0 0]{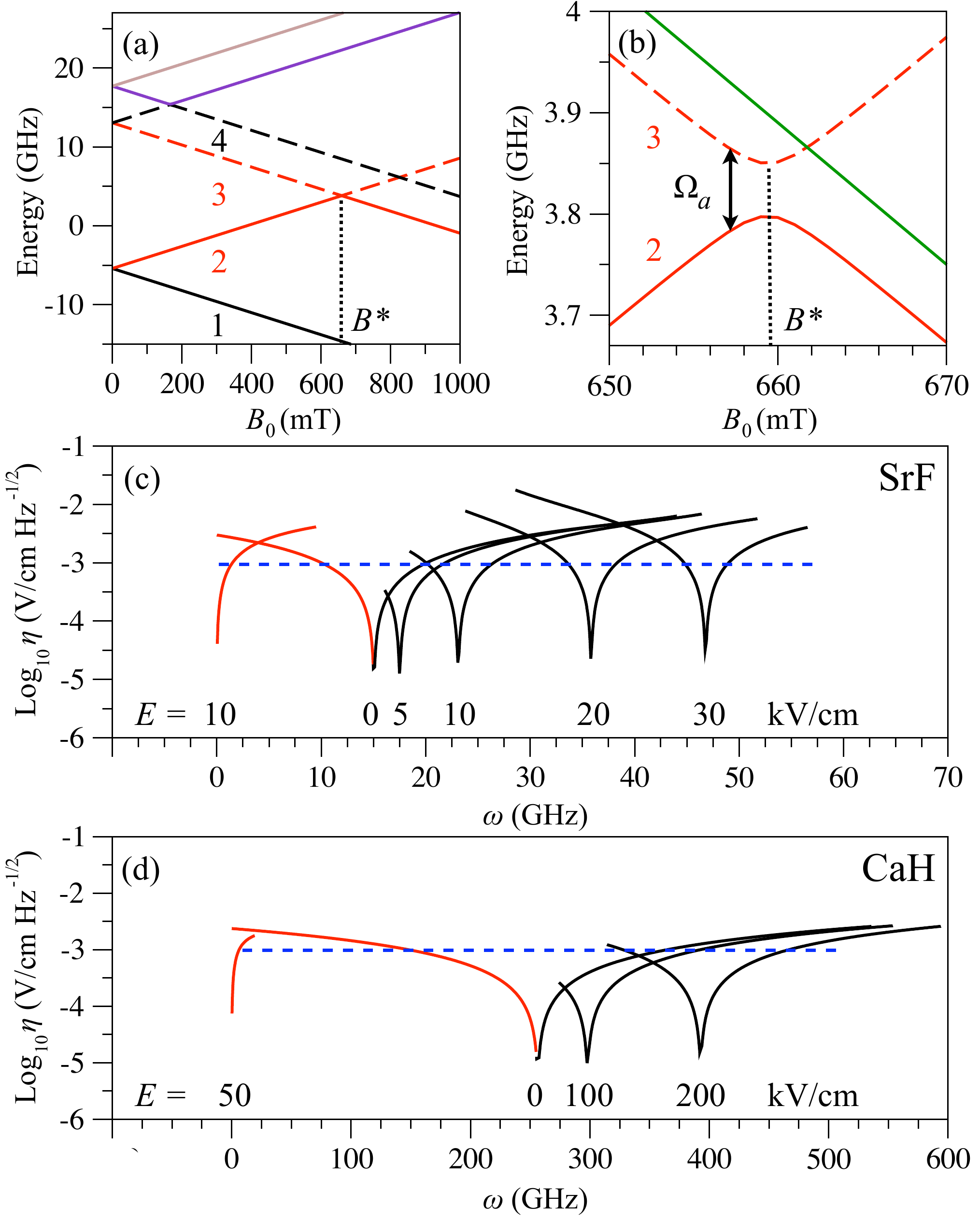}
\caption{\label{fig:SrFlevels} (a), (b) Energy levels of the SrF$(^2\Sigma^+)$ molecule ($b_{r}$ = 7.53 GHz, $\gamma$ = 74.7 MHz) in an electric field of $E_0$ = 10 kV/cm as a function of magnetic field $B_0$; (c) Frequency dependence of the ac field sensitivity for  SrF in a linearly polarized mw field for different electric fields; the red and black lines correspond to the $2\to3$ and $1\to4$ transitions respectively.  The dashed line represents the sensitivity to the magnetic field component of the ac field that can be achieved in experiments with atoms \cite{BohiAPL10}; (d) Same as in (c) but for the  CaH$(^2\Sigma^+)$ molecule ($b_{r}$ = 128.3 GHz, $\gamma$ = 1.24 GHz)}
\end{figure}

The effect of combined $\mathbf{B}_0$ and $\mathbf{E}_0$ fields on the rotational states of a $^2\Sigma$ molecule is shown in Fig.~\ref{fig:SrFlevels} (a) and (b) for the case of SrF ($^2 \Sigma^+$)~\cite{ShumanNature10,ShumanPRL09}.  We assume that the molecules are initially prepared in the magnetic low-field seeking $M$-component of the rotational ground state, state $\vert 2 \rangle \equiv \vert 0,  1/2, 1/2 \rangle$ in Fig.~\ref{fig:SrFlevels}(a). This can be achieved by cooling molecules to subKelvin temperatures and confining a  molecular cloud in a magnetic trap \cite{WeinsteinCaH}. Alternatively, molecules can be confined in an electric or optical trap and transferred to state $\vert 2 \rangle$ by a sequence of microwave pulses \cite{OspekausPRL10}. Molecules can be cooled by a variety of recently developed experimental techniques such as buffer-gas cooling, Stark deceleration, or laser cooling~\cite{WeinsteinCaH, BasPRL05, CampbellPRL07, ShumanNature10,ShumanPRL09}.

State $\vert 2 \rangle$ exhibits an avoided crossing with the magnetic high-field seeking state $\vert 3 \rangle \equiv  \vert 1,  1/2, 1/2 \rangle$ at the magnetic field $B^\ast = 659.5$ mT. 
%\begin{equation}
%\label{baster}
%	B^\ast = \frac{b_r}{g_S \mu_B} \left(1 - \frac{\gamma}{2 b_r} \right)  \left(2 + \frac{\gamma}{2 b_r} \right)
%\end{equation}
The states  $\vert 2 \rangle$ and $\vert 3 \rangle$ have the opposite parity and, due to the spin-rotation interaction, represent linear combinations of the states with spin projections $M_S= \pm 1/2$~\cite{FriHerPCCP00,TscherbulPRL06,PeresRiosNJP10}. Therefore the  $\vert 2 \rangle \to \vert 3 \rangle$ transition is dipole-allowed and can be used to detect the electric component of a resonant rf or microwave field.

\subsection{Detection of low-frequency ac fields}

The single mode electromagnetic field at point $r=0$ is given by
  \begin{equation}
\label{Efield}
	\mathbf{E}_\alpha (\mathbf{r},t) = E_\alpha (\mathbf{r})/2 \left[\mathbf{e_\alpha}(\mathbf{r}) \exp(- i\omega t) + \mathbf{e^\ast_\alpha} (\mathbf{r}) \exp( i\omega t)  \right],
\end{equation}  
 where $E_\alpha (\mathbf{r}),\mathbf{e}$ and $\omega$ are amplitude, polarization and frequency of the electromagnetic field. 
 The measurement requires placing the molecular ensemble close to the source of the field to be measured, and an addition of background dc  magnetic and electric fields, with magnitudes $\mathbf{B}_0$ and $\mathbf{E}_0$. 
 By adjusting $B_0$ and $E_0$, the energy splitting between the states $\vert 2 \rangle$ and $\vert 3 \rangle$ can be tuned in resonance with the field frequency $\omega$, as shown in Figure 1. The trapping field, if any, must be switched off in order to allow the field  $E (\mathbf{r},t)$ to drive resonant oscillations between the states $\vert 2 \rangle$ and $\vert 3 \rangle$ during the free evolution time $\tau$.  The  Rabi frequency of the oscillations is given by
\begin{equation}
\label{RabiF}
	\Omega_\alpha (B_0,E_0) = E_\alpha (\mathbf{r})  d_\alpha (B_0, E_0)/\hbar,
\end{equation}
where $\alpha=\{+;0; - \}$ denotes the polarization of the oscillating field with respect to the laboratory $Z$ axis, $E_\alpha (\mathbf{r})$ is the corresponding component of the field $E (\mathbf{r},t)$, and $d_\alpha (B_0, E_0)$ is the transition dipole moment between the two states. In order to compute $d_\alpha (B_0, E_0)$, we expand the molecular states $\vert n \rangle$ plotted in Fig.~\ref{fig:SrFlevels} as follows: 

\begin{equation}
|n \rangle= \sum_{N,M_N,M_S}c^{(n)}_{NM_NM_S}(B_0, E_0) | N M_N\rangle| S M_S\rangle, 
\label{opp}
\end{equation} 
which gives
 \begin{eqnarray}
d_\alpha (B_0, E_0) = \sqrt{\frac{4 \pi}{3}}  \sum_{N,M_N,M_S} \sum_{N',M_N',M_S'} c^{(2)\ast}_{NM_NM_S}c^{(3)}_{N'M'_NM'_S}
\nonumber
\\
\langle N M_N|Y_{1 \alpha} | N' M_N' \rangle \delta_{M_S,M_{S}'}.
\hspace{1.cm}
\label{AOz}
\end{eqnarray} 
Here, $M_N$ and $M_S$ denote the projections of the angular momenta ${\bf N}$ and ${\bf S}$ on the $Z$ axis, respectively.

%This crossing is allowed in a magnetic field, but becomes avoided once an additional electric field is applied.

%\textit{Measuring microwave fields}

%Measuring oscillating fields with a Ramsey-type experiment is not an easy task since one need to synchronize the spin-echo pulse sequence with ac field oscillations~\cite{TaylorNatPhys08}.

\begin{figure}[t]
\includegraphics[width=7.cm, trim = 30 0 0 0]{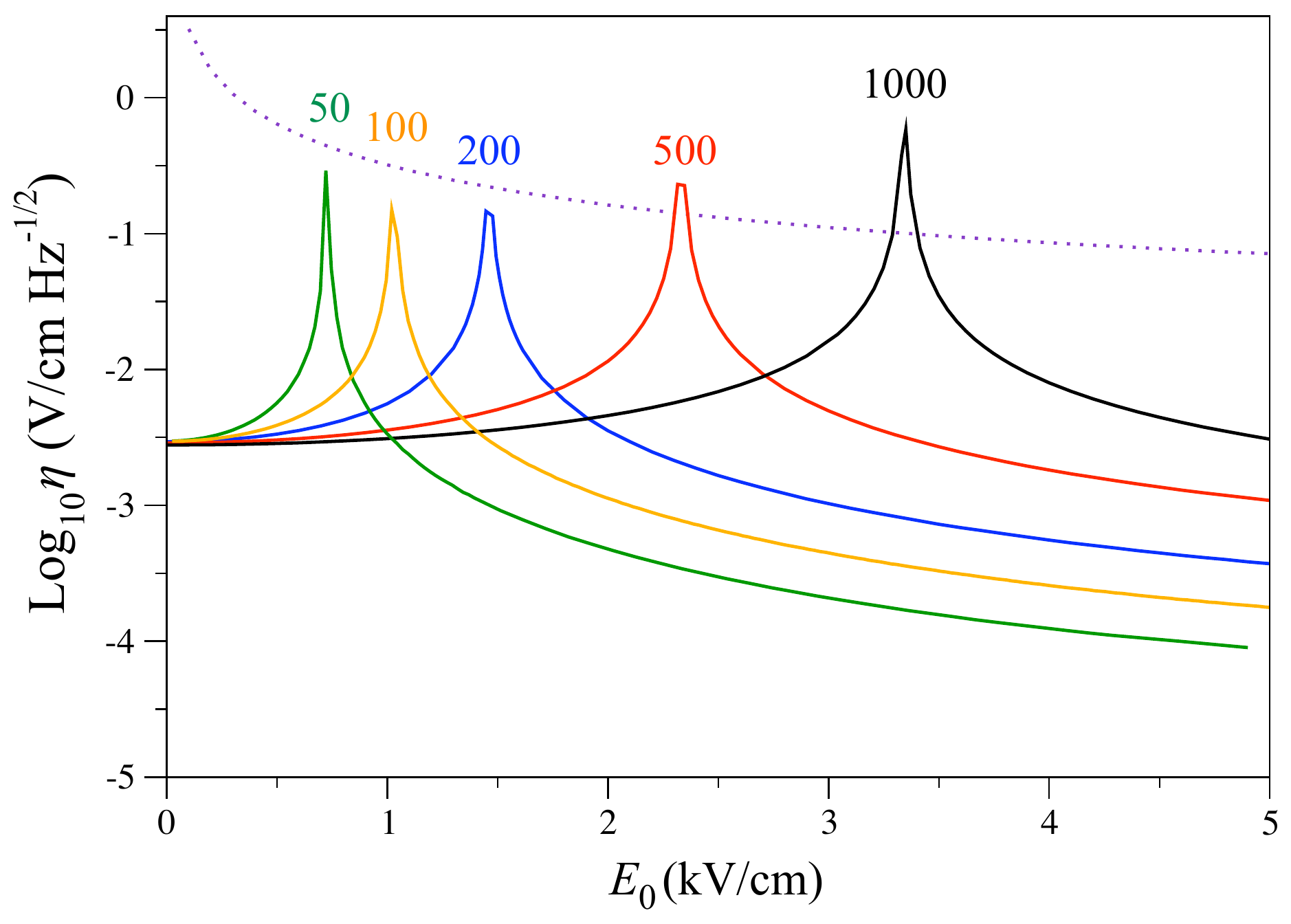}
\caption{\label{fig:ef-dependence} Sensitivity of the $|2\rangle \rightarrow |3 \rangle$ transition in SrF$(^2\Sigma^+)$ to electromagnetic fields as a function of the dc electric field $E_0$ and the frequency of the microwave field. The peaks are labeled by the frequency of the ac field in MHz. The dotted line shows the sensitivity of the $|2\rangle \rightarrow |3 \rangle$ transition in SrF$(^2\Sigma^+)$ to the magnetic field component of the microwave field.}
\end{figure}

 \begin{figure}[t]
	\centering
	\includegraphics[width=8cm, trim = 0 0 0 0]{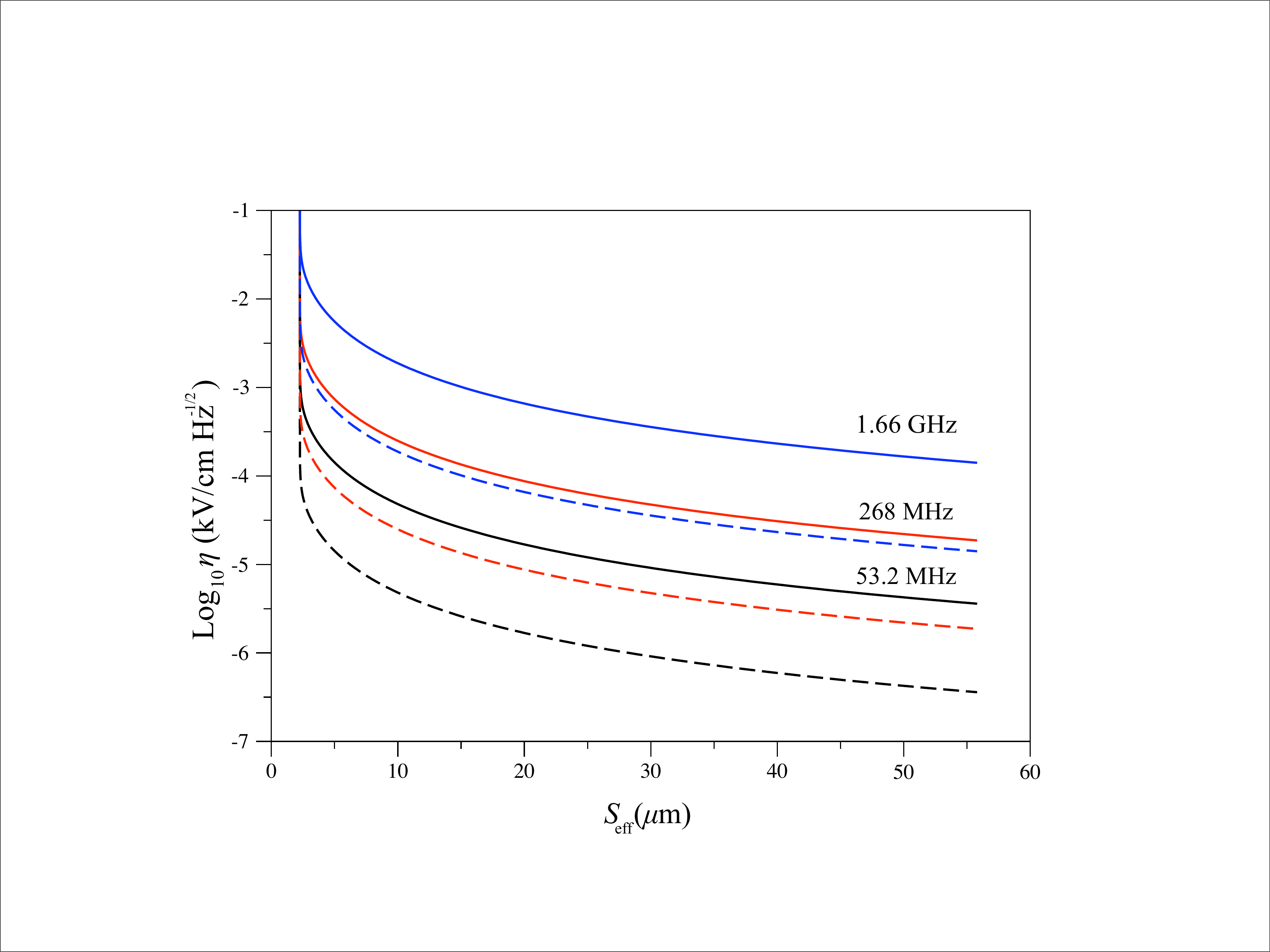}
	\caption{\label{Sensitiv2}
	The ac electric field sensitivity for SrF$(^2\Sigma^+)$ as a function of the spatial resolution, for $n=10^{10}$~cm$^{-3}$ (solid lines) and $n=10^{12}$~cm$^{-3}$  (dashed lines). Different colors correspond to different ac field frequencies.} 	
\end{figure}
After time $\tau$ the probability to detect a molecule in the state $\vert 3 \rangle$ at the spatial point $\mathbf{r}$ is
\begin{equation}
\label{Prob2}
	p_3 (\mathbf{r}) = \frac{n_3(\mathbf{r})}{n_2(\mathbf{r})+n_3(\mathbf{r})} = \sin^2 \left[ \frac{\Omega_\alpha (\mathbf{r}) \tau}{2} \right],
\end{equation}
where $n_{2}$ and $n_3$ are the densities of the molecules in the states $\vert 2 \rangle$ and $\vert 3 \rangle$. 
The detection of the population of state $\vert 3 \rangle$ can be done, for example, by using the direct absorption imaging technique~\cite{JunYePRA10} or resonance enhanced multi photon ionization technique (REMPI \cite{rempi}), allowing one to measure the population in a single-shot experiment.
From the measured value of $p_3(\mathbf{r})$ one can calculate $\Omega_{\alpha}(\mathbf{r})$ and, using Eq.~(\ref{RabiF}),  the components of the electric field. The amplitudes, $E_{x}(\mathbf{r})$, $E_{y}(\mathbf{r})$, and $E_{z}(\mathbf{r})$, and the relative phases can be reconstructed by measuring $E_{\alpha}(\mathbf{r})$ with the background magnetic field $\mathbf{B}_{0}$ pointing along the $x$, $y$, and $z$ axes, by analogy with ref. \cite{BohiAPL10}. From $\mathbf{E} (\mathbf{r})$, the spatial distribution of the magnetic field  can be calculated using the Maxwell equations, while the amplitudes of the two fields are related as $E=c B$.

\subsection{Sensitivity and spatial resolution}

The single-shot sensitivity of the measurement to the ac electric fields is given by~\cite{BohiAPL10}:
\begin{equation}
\eta^\text{ac}_{E} \text{[V/cm Hz$^{-1/2}$]} = \frac{2\sqrt3\hbar}{100~d_\alpha (B_0, E_0) \sqrt{nV_\text{eff}}\sqrt{\tau}} 
\label{Eacsensitivity}
\end{equation} 
where $n$ is the density of molecules, $V_\text{eff}= 2 \pi \sigma_\text{eff}^2 \rho$ is the effective imaging volume, $\sigma_\text{eff}$ is the dispersion of the spatial coordinate, and $\rho$ represents the $1/e$ radius of the cloud.

The dependence of the transition dipole moment $d_\alpha$ on the background fields $B_0$ and $E_0$ renders the sensitivity frequency dependent, $\eta^\text{ac}_{E} \equiv \eta^\text{ac}_{E}(\omega)$, as shown  in Fig.~\ref{fig:SrFlevels}~(c)  by red curves  for the $\vert 2 \rangle \to \vert 3 \rangle$ transition in SrF$(^2\Sigma^+)$. One can see that the frequency dependence exhibits sharp minima reaching the sensitivities on the order of $10^{-6} - 10^{-7}$ V/cm~Hz$^{-1/2}$. The positions of the minima can be controlled by tuning the splitting of the levels $\vert 2 \rangle$ and  $\vert 3 \rangle$ with the background electrostatic field $E_0$, and thereby shifted towards smaller frequencies~\footnote{The lowest detectable ac field frequency, $\omega_\text{min}$, is limited by the linewidths of the dipole-dipole broadening, $\Gamma = 8/9\pi d^{2}n$, and the Doppler broadening, $\sigma_{\omega}=\sqrt{kT/mc^2}\omega$, of the $\left | 2 \right > \to \left | 3 \right >$ transition. For a gas of SrF molecules with a density $n=10^{10}$~cm$^{-3}$ the value of  $\omega_\text{min}$ ranges from 0.26 kHz at $T=1~\mu$K to 8.28~kHz at $T=1
$~K.}. The accessible frequency range can be extended by initially preparing the SrF molecules in the high-field seeking state,  $\vert 1 \rangle \equiv \vert 0, 1/2, -1/2 \rangle$, and driving the transition to the state  $\vert 4\rangle \equiv \vert 1, 3/2, -3/2 \rangle$. The frequency dependence of the sensitivity corresponding to the $\vert 1 \rangle \to \vert 4 \rangle$ transition is shown in Fig.~\ref{fig:SrFlevels}~(c) by black lines. Choosing a molecule with a different rotational constant allows for the detection of a completely different range of accessible frequencies. As an example, the rotational constant of CaH$(^2\Sigma^+)$ molecule~\cite{WeinsteinCaH} is about 17 times larger than that of SrF, which gives access to  microwave fields of frequencies $\omega \sim 100- 500$ GHz, as shown in Fig.~\ref{fig:SrFlevels}~(d).

Interestingly, the transition dipole moment $d_\alpha (B_0, E_0)$ for the $|2\rangle \rightarrow |3 \rangle$ transition in $^2\Sigma$ molecules vanishes at certain combinations of $E_0$ and $B_0$. At these particular combinations, the molecules become transparent to the resonant microwave field. This is demonstrated in Figure~\ref{fig:ef-dependence}. The figure illustrates that the magnitude of the dc electric field $E_0$, for which $d_\alpha (B_0, E_0) \sim 0$, depends sensitively on the frequency of the resonant transition (which can be tuned by varying $B_0$). This can be used for sensitive detection of the magnitude $E_0$ of a dc electric field, given the magnitude of $B_0$ and the frequency of the microwave field. Conversely, this can be also used for sensitive detection of the magnitude $B_0$ of the dc magnetic field, given  
the magnitude of $E_0$ and the resonant microwave frequency.

Longer interaction time $\tau$ results in increased sensitivity to electric fields. The sensitivity is, however, gained at the expense of the spatial resolution  that decreases with $\tau$ due to the molecular motion and decoherence. The effective spatial resolution, $S_\text{eff}=2 {(\sigma_{\tau}^2+\sigma_{\rm img}^2+\sigma_{\rm ps}^2)}^{1/2}$, can be calculated from the displacements  $\sigma_\tau = \tau \sqrt{2k_{B}T/m}$ during the measurement time ($\tau$), $\sigma_\text{img} = \tau_\text{img} \sqrt{2k_{B}T/m}$ during the imaging pulse ($\tau_\text{img}$), and $\sigma_{\rm ps} = v_\text{rec} \tau_\text{img} \sqrt{2N_p}/3$ due to photon scattering. Here $k_B$ is  the Boltzmann constant, $T$ is the temperature of the gas, $m$ is the mass of the molecule, $v_\text{rec} = 5.6 {\rm mm/s}$ is the recoil velocity, $N_p=1/(1-\eta)$ is a number of photons scattered by each molecule before it goes to a dark state, and $\eta$ is a branching factor equal to Franck-Condon factor~\cite{ShumanNature10,ShumanPRL09} multiplied by the H{\"o}nl-London factor. 
To our knowledge, the H{\"o}nl-London factor for SrF molecules is not available in the literature, therefore for our estimates we use the value 2/3 that was determined for KRb molecules in an experiment reporting the absorption imaging of ultracold molecules~\cite{JunYePRA10}. 
The average displacement due to photon scattering durring imaging time $\tau_\text{img} = 40 \mu s$ is $\sigma_{\rm img} = 0.2 \mu m$. Fig.~\ref{Sensitiv2} illustrates the relation between the sensitivity and the spatial resolution for an ensemble of SrF molecules and different frequencies of the field detected.

\begin{figure}[t]
\includegraphics[width=9.cm, trim = 30 0 0 0]{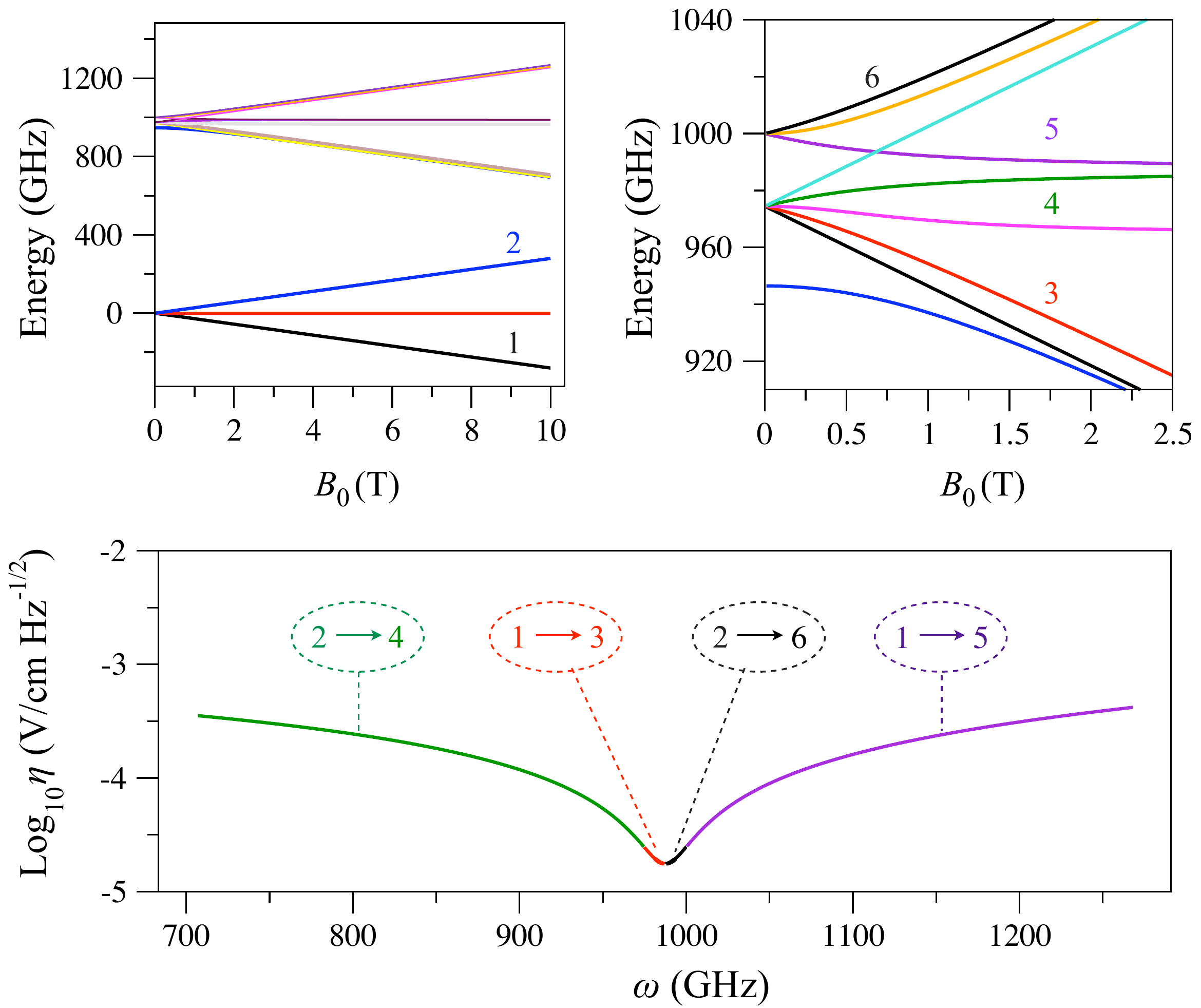}
\caption{\label{fig:NHlevels} (a), (b) Energy levels of the NH$(^3\Sigma^-)$ molecule ($b_{r}$ = 490.0 GHz, $\gamma = - 1.65$ GHz, and $\lambda=27.6$ GHz) as a function of the magnetic field $B_0$; (c) Frequency dependence of the ac field sensitivity. The background electric field  $E_0=0$.}
\end{figure}

The energy level structure of paramagnetic molecules can also be used to probe sensitively static or off-resonant rf and microwave fields. This can be achieved by measuring the phase accummulation in a Ramsey-type sequence consisting of two $\pi/2$ pulses~\cite{RamseyMolBeams}. The first $\pi/2$ pulse prepares the molecules in the equal superposition of states $\vert 2 \rangle$ and $\vert 3 \rangle$, which acquire a relative phase proportional to the Stark shift, 
$\Delta \phi= d_{\rm{eff}}(B_0,E_0) E \tau$,
due to the effective dipole moment $d_{\rm{eff}}(B_0,E_0) = \frac{d(\hbar \omega_{32})}{dE}$ during the evolution time $\tau$.
The second $\pi/2$-pulse transforms the relative phase into a population difference. 
Because the states $|2 \rangle$ and $| 3 \rangle$ become nearly degenerate at magnetic fields near $B = B^\ast$, the magnitude of $d_{\rm{eff}}$ is significatly enhanced near the avoided crossing depicted in Figure 1 (b). The sensitivity to a dc electric field  is given by
\begin{equation}
\label{Sens}
\eta^{\rm{dc}}_{\rm{E}} = \frac{2\sqrt{3}\hbar}{100
 d_{\rm{eff}}(B_0,E_0)\sqrt{n V_{\rm{eff}}}\sqrt{\tau}},
\end{equation}
and equals $1.58 \times 10^{-5}$ V/cm Hz$^{-1/2}$ for SrF molecules 
with the density $10^{12}$ cm$^{-3}$ in a magnetic field of  $B_0 = 545$ mT and electric field of  $E_0 = 2.5$ kV/cm.

The range of detectable frequencies of electromagnetic fields can be extended by using paramagnetic molecules of higher spin multiplicity. For example, $^3\Sigma$ molecules offer a series of tunable transitions that can be used to probe ac electric fields in the same way as the transitions in the $^2\Sigma$ molecules described above. This is illustrated in Fig.~\ref{fig:NHlevels} that shows the energy level structure of NH$(^3\Sigma^-)$ molecules, and the corresponding sensitivities. A large value of the rotational constant of NH and a series of tunable dipole-allowed transitions  allow for the possibility to cover continuously a broad range of detectable ac fields in the THz frequency region, which is particularly interesting for a variety of practical applications \cite{MittlemanRPP97}.

The sensitivity to the magnetic component of an ac field can be calculated as $\eta^\text{ac}_{B} \text{[T Hz$^{-1/2}$]} = \eta^\text{ac}_{E} \text{[V/cm Hz$^{-1/2}$]} \times 3.336\cdot 10^{-7}$. The ratio of the minimal detectable magnetic fields in experiments with cold atoms~\cite{BohiAPL10}  and molecules is given by $B^\text{at}_\text{min}/ B^\text{mol}_\text{min}= 2 c  d^{12}_\alpha/(\sqrt{3}\mu_{B})$(assuming the same interaction time $\tau$, the same number of particles and the same detection efficiency in  both cases), which for typical molecules with $d\sim1$ a.u. amounts to $\sim$100-fold higher sensitivity using the proposed scheme.

%  \begin{figure}
%	\centering
%	\includegraphics[width=0.5\textwidth, trim = 0 40 0 40]{fig2n}
%	\label{Sensitiv}
%	\caption{
%	D.C and A.C. electric field sensitivities for gas of SrF molecules (of densities 2.2*10$^{11}$cm$^{-3}$ - solid line and 4.7*10$^{15}$cm$^{-3}$ - dashed line) as a function of applied magnetic field \textbf{[to the suplementary]}
%} 	
%\end{figure}

%\vspace{1cm}

%\vspace{1cm}

%\textit{Conclusions}

\section{Summary}

In summary,  we have described a technique for sensitive parallel measurements of electric and magnetic field components of electromagnetic fields, both dc and oscillating with frequencies ranging from a fraction of a kHz to THz. The method, based on tunable energy level structure of paramagnetic molecules in superimposed electric and magnetic fields, allows one to achieve the sensitivity on the order of $\mu$V/cm~Hz$^{-1/2}$ and 100 fT~Hz$^{-1/2}$  for the ac fields and $\sim 10$ $\mu$V/cm~Hz$^{-1/2}$ and nT~Hz$^{-1/2}$  for dc fields. The sensitivity of the technique can be further enhanced by employing the spin-echo pulse sequence~\cite{HahnPR50}, which can be used e.g. to characterize the field of a microwave stripline or map out the spatial distribution of electron spins. Finally, the method proposed here can be used for detecting weak dc and ac fields in the presence of high backround dc magnetic and electric field.

The work was supported by NSERC of Canada and the NSF grant to ITAMP at Harvard University and the Smithsonian Astrophysical Observatory.

\bibliography{References_library}

\end{document}